\documentclass{pasj01}
\draft 
\Received{$\langle$reception date$\rangle$}
\Accepted{$\langle$acception date$\rangle$}
\Published{$\langle$publication date$\rangle$}

\begin{document}

\title{Physical properties of the molecular cloud, N4, in SS433; Evidence for an interaction of molecular cloud with the jet from SS433}

\author{Hiroaki Yamamoto$^1$, Ryuji Okamoto$^1$, Yasuhiro Murata$^2$,  Hiroyuki Nakanishi$^3$, Hiroshi Imai$^{3,4}$, and Kohei Kurahara$^5$}%
\altaffiltext{1}{Graduate School of Science, Nagoya University, Furo-cho, Chikusa-ku, Nagoya, Aichi 464-8602, Japan}
\altaffiltext{2}{Institute of Space and Astronautical Science, Japan Aerospace Exploration Agency, 3-1-1, Yoshinodai, Chuo-ku, Sagamihara, Kanagawa 252-5210, Japan}
\altaffiltext{3}{Amanogawa Galaxy Astronomy Research Center, Graduate School of Science and Engineering, Kagoshima University,  \\
1-21-35 Korimoto, Kagoshima 890-0065, Japan}
\altaffiltext{4}{Center for General Education, Institute for Comprehensive Education, Kagoshima University,  \\
1-21-30 Korimoto, Kagoshima 890-0065, Japan}
\altaffiltext{5}{Mizusawa VLBI Observatory, National Astronomical Observatory of Japan, 2-21-1, Osawa, Mitaka, Tokyo 181-8588, Japan}

\email{hiro@a.phys.nagoya-u.ac.jp}


\maketitle

\begin{abstract}
We conducted observations and analyses of the molecular cloud, N4, which is located at $\sim$40 pc from SS433 and the same line of sight as that of the radio shell, in $^{12}$CO($J$=1--0), $^{12}$CO($J$=3--2), $^{13}$CO($J$=3--2), and grand-state OH emissions. 
N4 has a strong gradient of the integrated intensity of $^{12}$CO($J$=1--0, 3--2) emission at the northern, eastern and western edges. 
The main body of N4 also has a velocity gradient of $\sim$0.16 km s$^{-1}$ (20$\arcsec$)$^{-1}$.
A velocity shift by up to 3 km s$^{-1}$ from the systemic velocity at $\sim$49 km s$^{-1}$ is detected at only the northwestern part of N4. 
The volume density of the molecular hydrogen gas and the kinematic temperature are estimated at eight local peaks of $^{12}$CO($J$=1--0) and $^{13}$CO($J$=3-2) emissions by the RADEX code. 
The calculated $n_{\rm (H_2)}$ is an order of 10$^3$ cm$^{-3}$, and $T_{\rm k}$ ranges $\sim$20 K to $\sim$56 K. 
The mass of N4 is estimated to be $\sim$7300 $M_{\rm \odot}$. 
The thermal and turbulent pressures in N4 are estimated to be $\sim$10$^5$ K cm$^{-3}$ and $\sim$10$^7$ K cm$^{-3}$, respectively. 
The relation of the thermal and turbulent pressures in N4 tends to be similar to that of the molecular clouds in the Galactic plane. However, these values are higher than those in the typical molecular clouds in the Galactic plane. 
Several pieces of circumstantial evidence representing the physical properties of N4 and comparison with the data of infrared and X-ray radiation suggest that N4 is interacting with a jet from SS433. However, no gamma-ray radiation is detected toward N4.
Compared to the previous study, it is hard to detect the gamma-ray radiation by cosmic-ray proton origin due to the low sensitivity of the current gamma-ray observatories.
Any OH emission was not detected toward N4 due to the low sensitivity of the observation and antenna beam dilution.
\end{abstract}


\section{Introduction}

The microquasar SS433 has been well known as an X-ray binary located at ($\alpha$, $\delta$)(J2000) $\sim$ (\timeform{19h11m49.6s}, \timeform{4D58'57.8''}) which is the nearly center of the supernova remnant W50 detected in the radio continuum \citep{dub98}. 
SS433 emanates a powerful jet whose velocity is $\sim$0.26$c$ \citep{mar84}. 
The jet has a spiral structure with a half opening angle of $\sim$20 degrees, and the precession period of the jet is estimated to be $\sim$162.5 d (e.g., \cite{mar79}; \cite{mar84}; \cite{blu04}). 
The radio shell is elongated in the east--west direction due to the bursting of the powerful jet. Possible interaction of the SS433/W50 system with an ambient interstellar medium (ISM) has been discussed (e.g., \cite{dub98}; \cite{loc07}; \cite{yam08}; \cite{su018}; \cite{liu20}; \cite{sak21}). 

Several observations and analyses have been attempted to determine the distance to SS433. 
\citet{ver93} estimated it to be 4.85$\pm$0.02 kpc from the proper motions of the blobs ejected to both sides from SS433 by monitoring for several days in VLBI. 
\citet{dub98} estimated it using the spatial correlation between the morphologies of radio continuum emission and the H$_{\rm I}$ emission at $V_{\rm LSR}$ of $\sim$42 km s$^{-1}$, corresponding to a kinematic distance of 3 kpc derived from the Galactic rotation model (e.g., \cite{deh98}), where $V_{\rm LSR}$ is relative velocity from the local standard of rest defined as mean motion in the neighborhood of the Sun.  
\citet{sak21} also compared the radio shell with GALFA-HI survey datasets taken with Arecibo 305 m with a spatial resolution of 4 arc-min  \citep{pee18}.
They found that good anti-correlation between the radio shell and the H$_{\rm I}$ distribution is seen at a velocity of $\sim$40 km s$^{-1}$, corresponding to a distance of 3 kpc.
The distance to SS433 has also been estimated to be 5.5$\pm$0.2 kpc by applying a deep-integration radio image of SS433 on the arcsecond scale to the kinematic model based on the speed and the precession period of the jet (\cite{blu04}; \cite{loc07}). 
Gaia determined the parallax of the companion star of SS433 seen in visible light or optical wavelength. 
The distance estimated to be $\sim$4.6 kpc with Data Release 2 \citep{gai18} was revised to 8.5 kpc with Early Data Release 3 \citep{gai21}. 
These mean that the distance to SS433 still has uncertainty even in recent estimation. 
In this paper, we adopt the distance to SS433 to be 5.5 kpc. 

X-ray observations revealed the spatial distribution and properties of the jet emanating from SS433 (e.g., \cite{yam94}; \cite{saf97}; \cite{kot98}; \cite{mol05}; \cite{bri07}). 
The X-ray jet is extended by $\sim$0.7 degrees and  $\sim$1 degree at the western and eastern sides, respectively, corresponding to $\sim$70 pc and $\sim$100 pc at a distance of 5.5 kpc. 
The kinetic power and the mass outflow rate of the jet were estimated to be $\sim$3.2 $\times$ 10$^{38}$ ergs s$^{-1}$ and 1.5 $\times$ 10$^{-7}$ M$_{\rm \odot}$ yr$^{-1}$, respectively, by assuming the distance of 4.85 kpc estimated in \citet{ver93} \citep{mar02}. 
The dense ISM in and around the SS433/W50 system can be a site of the gamma-ray radiation induced by an interaction of a high energy cosmic-ray proton accelerated through diffusive shock acceleration \citep{fer49} with a hydrogen atom (hadronic process). 
High energy gamma-ray radiation at a few hundred GeV to 100 TeV has been detected by Fermi-LAT, HAWC (High Altitude Water Cherenkov Observatory) (\cite{abe18}; \cite{fan20}).  
The origin of such GeV-TeV gamma-ray radiation is considered to be inverse Compton scattering. 
\citet{liu20} discussed the possibility to detect gamma-ray radiation by the hadronic process but concluded that the sensitivity of the current gamma-ray observatories is too low to detect such gamma-ray radiation. 
The first Large High Altitude Air Shower Observatory (LHAASO) observations covering the northern hemisphere resulted in no detections of ultra high energy gamma-ray in the SS433/W50 system.
This may be due to current short photon-collecting time or intrinsically low gamma-ray emissivity at the energy band etc.

Numerical simulations have been conducted to understand the nature of the SS433/W50 system (e.g., \cite{asa14}; \cite{sud20}; \cite{ohm21}). 
The interaction of the jet with interstellar medium (ISM) was discussed by magnetohydrodynamical (MHD) simulations \citep{asa14} and their simulation indicated that such an interaction can make dense ISM.

The main lines of the grand-state of hyperfine structure in OH (1665, 1667MHz) could be a better tracer of molecular gas around an envelope of molecular clouds traced in CO emission and lower density molecular clouds not traced in CO emission called as CO-dark gas (e.g., \cite{bar10}; \cite{all15}; \cite{ngu18}; \cite{don19}). 
The numerical simulation also predicted that an abundance of OH is higher than that of CO in a low $A_{\rm V}$ condition, suggesting that the OH line could be a good tracer of the low-density molecular clouds (e.g., \cite{wol10}). 
The satellite lines of OH (1612, 1720MHz) are also crucial as a shock tracer. 
In a specific density and temperature, the state population is inverted, and the line can often be detected as maser (e.g., \cite{loc99}). 

$^{12}$CO($J$=1--0) observations toward SS433 have been first conducted by \citet{hua83}. 
They concluded that SS433 is associated with the edge of the large filament at 27--36 km s$^{-1}$, while recent results suggest that the large filament is located at $\sim$1.7 kpc far from the SS433/W50 system \citep{lin20}. 
\citet{dur00} detected $^{12}$CO emission at Blob 4 and 5 in the western part of SS433 at $\sim$50 km s$^{-1}$, where Blob 5 is nearly the same direction as N4 \citep{yam08}. 
Fifteen $\mu$m radiation is also detected with ISOCAM at the molecular clouds \citep{fuc01}. 
The molecular clouds detected in CO emission and associated with the SS433/W50 system were identified by \citet{yam08} using the NANTEN Galactic Plane Survey dataset \citep{miz04}. 
They found ten molecular clouds aligned along the jet axis.
Four out of the ten clouds are located at the western part of the SS433/W50 system and others at its eastern part but far from the radio continuum shell. 
The LSR velocity of the molecular clouds is $\sim$49 to $\sim$56 km s$^{-1}$ at the western side and $\sim$42 to $\sim$45 km s$^{-1}$ at the eastern side, corresponding to a kinematic distance of 3--3.5 kpc when applying the Galactic rotation model \citep{deh98}. 
The molecular clouds at the western side named as N1, N2, N3, and N4 in \citet{yam08} have good spatial correlation/anti-correlation with the radio continuum shell.
\citet{su018} discussed the association of the SS433/W50 system with the molecular clouds at LSR velocity of $\sim$80 km s$^{-1}$, corresponding to $\sim$5 kpc.
\citet{liu20} have conducted a detailed analysis of the molecular clouds in space and velocity toward N2 and N3 with the IRAM 30m telescope in $^{12}$CO($J$=1--0, 2--1) and CN($J$=3/2--1/2). 
They concluded that the molecular clouds are associated with the SS433/W50 system. However, their observations have not covered N4. 

N4 is an important target because it is located at the same line of sight as the radio shell while others are located at the edge of the radio shell.
In this paper, first we present the whole distribution of the molecular clouds, N1--N4, named in \citet{yam08}, detected with Nobeyama 45 m telescope (NRO45m) toward the western part of the SS433/W50 system. 
Then we focus on N4 and present a detailed analysis with the data of $^{12}$CO($J$=3--2) and $^{13}$CO($J$=3--2) emission taken with James Clark Maxwell Telescope (JCMT15m).
We also present the results of our observations of OH emission to search for possible thermal or maser emissions to understand the nature of N4.

\section{Observations}

\subsection{Nobeyama 45m telescope}
Observations were conducted toward the molecular clouds detected by NANTEN observations in \citet{yam08}. 
Total 33 hours in the observations were taken in several sessions between March to May in 2010 with NRO45m. 
In this paper, the dataset at only the western part of SS433 is used and other papers will publish the dataset at the eastern part. 
BEARS (25-BEam Array Receiver System) was used as a front-end system \citep{sun00} for the observations in $^{12}$CO($J$=1--0) emission whose rest frequency was 115.27120 GHz. 
The beam size of the telescope at 115 GHz was $\sim$15$\arcsec$. 
The backend was Auto Correlators, and 32 MHz bandwidth mode was used for 25 beams \citep{sor00}. 
Velocity resolution was $\sim$0.1 km s$^{-1}$. 
$T_{\rm sys}$ was 200--400 K, where $T_{\rm sys}$ is the temperature obtained by converting the noise power originated from the observing system and the atmosphere into temperature. 
On-The-Fly mapping mode was applied in the observations by scanning directions of the Galactic longitude and latitude (\cite{man00}; \cite{saw08}). 
Orthogonal two scans were convolved by the basket weaving method \citep{eme88}. 
The observing grid with the OTF mapping mode was 5$\arcsec$, and an output data grid was 7$\arcsec$.2 which is the same grid size as the JCMT archival data.
The data was smoothed to $\sim$20$\arcsec$ to reduce the r.m.s. noise levels. 
The final r.m.s. noise level of the data was $\sim$1.4 K ch$^{-1}$.  
The three-sigma level of the data was 1.9 K km s$^{-1}$. 
The pointing accuracy has been kept within 3$\arcsec$ by observing radio point sources listed by the observatory every two hours. 
The intensity calibration was conducted by S100 receiver by observing W51.

\subsection{JCMT archival data}
Archival datasets taken with the JCMT15m were used. 
The proposal ID of the data used in this study is M18BP027. 
The western side of SS433 was covered in $^{12}$CO, $^{13}$CO, C$^{18}$O($J$=3--2), and H$^{13}$CO$^{+}$($J$=4--3). 
The frequency resolution of the data was 61 kHz, corresponding to the velocity resolution of $\sim$0.05 km s$^{-1}$. The velocity resolution was smoothed to 0.1 km s$^{-1}$  in this paper. 
The spatial grid of the data was 7\arcsec.2 for the beam size of $\sim$15\arcsec.
The data was smoothed to 20$\arcsec$ to be compared with $^{12}$CO($J$=1--0).
$\eta_{\rm MB} = $0.72 was adopted as the main beam efficiency \citep{buc09} to convert the $T_{\rm a}$, the intensity scale in the archival data to the main beam temperature, $T_{\rm MB}$, scale. 
The final r.m.s. noise level of the data in $^{12}$CO, $^{13}$CO, C$^{18}$O($J$=3--2), and H$^{13}$CO$^{+}$($J$=4--3) was $\sim$1.6, $\sim$1.9, $\sim$2.4, and $\sim$1.4 K ch$^{-1}$, respectively. 
The three-sigma level of the data in $^{12}$CO, $^{13}$CO, C$^{18}$O($J$=3--2), and H$^{13}$CO$^{+}$($J$=4--3) were $\sim$2.6, $\sim$2.5, $\sim$3.2, and $\sim$1.9 K km s$^{-1}$, respectively. 
In this paper, only the datasets toward N4 are used. 
Since C$^{18}$O($J$=3--2) and H$^{13}$CO$^{+}$($J$=4--3) were not detected toward N4, the detailed analysis for only $^{12}$CO and $^{13}$CO data was conducted.

\subsection{Usuda 64m telescope}
The Usuda 64 m telescope, which is usually used for communication with deep space exploration operated by Japan Aerospace eXploration Agency (JAXA), was used for the observations in the ground-state of an OH emission in $J$=3/2, $F$=1--2, 1--1, 2--2 and 2--1 whose rest frequencies are 1612.23101, 1665.40184, 1667.35903, and 1720.52988 MHz, respectively. 
The observation was carried out on 3rd December 2019. 
The four lines were observed simultaneously by the frequency switching mode whose switching time and frequency intervals are 30 seconds and 500 kHz, respectively. 
The K5/VSSP, K5/VSSP32, and ADS3000+ were used as the backend. 
In this paper, only ADS3000+ was used to conduct a detailed analysis. 
The mode of 4 MHz bandwidth with 8192 channels was applied to the observations. 
The frequency resolution of the data was 488 Hz, corresponding to a velocity resolution of $\sim$0.087 km s$^{-1}$ at 1667 MHz. 
The beam size of the telescope at 1665 MHz is $\sim$10\arcmin.7. 
Only a single pointing observation at ($\alpha$, $\delta$)($J$2000) $\sim$ (\timeform{19h10m14.792s}, \timeform{5D4'52.504''}) was conducted in total 30 min. 
The final r.m.s. noise level of the data was $\sim$100 mK in $T_{\rm a}$ scale.

\section{Results}

\subsection{Spatial distribution of CO emission}
Figure 1 shows the overall CO intensity distribution toward the western part of SS433. 
Molecular clouds are well associated with the 1.4 GHz radio continuum source found with the VLA \citep{dub98}. 
The molecular clouds, N1, N2, and N3, located at the edge of the radio continuum shell, have also been studied in $^{12}$CO($J$=1--0, 2--1) with the IRAM 30 m telescope \citep{liu20}. 
It is concluded that these molecular clouds are associated with SS433. 
In this paper, the molecular cloud located at ($\alpha$, $\delta$)($J$2000) $\sim$ (\timeform{19h10m15s}, \timeform{5D4'0''}) is focused. 
Figure 2 shows the integrated intensity maps of (a) $^{12}$CO($J$=1--0), (b) $^{12}$CO($J$=3--2) and (c) $^{13}$CO($J$=3--2) in N4. 
Molecular cloud in $^{12}$CO looks like a triangle, and has bright spots at northeastern and southwestern parts.
The integrated intensity of $^{12}$CO($J$=1--0) reaches $\sim$60 K km s$^{-1}$, corresponding to 1.2$\times$10$^{22}$ cm$^{-2}$ if assuming the X-factor of 2$\times$10$^{20}$ cm$^{-2}$ (K km s$^{-1}$)$^{-1}$ \citep{bol13}. 
The intensity gradient at the northern, western, and eastern edge of N4 is steep, and the gradient of N4 at the northern edge is as high as $\sim$11.5 K km$^{-1}$-1 (20$\arcsec$)$^{-1}$.
The distribution of $^{12}$CO($J$=3--2) emission are similar to that of $^{12}$CO($J$=1--0). 
$^{13}$CO($J$=3--2) is detected in the region where $^{12}$CO emission is intense.
The integrated intensity in $^{13}$CO($J$=3--2) has a more clumpy structure, and its intensity distribution is different from that in $^{12}$CO emission.
In the $^{12}$CO emission, the peak at the eastern side is stronger than that at the western side. 
However, in the $^{13}$CO emission, the peak at the western side is stronger than that at the eastern side. 

\subsection{Velocity structure}
Figure 3 shows the position-velocity diagram of the $^{12}$CO($J$=1--0) emission integrating along the Right Ascension. 
The regions of the data used in figures 3(a) and 3(b) are shown by solid boxes in figure 2(a). 
The velocity of the eastern part of N4 ranges from 47.5 to 51 km s$^{-1}$. 
The velocity gradient of $\sim $0.16 km s$^{-1}$ (20$\arcsec$)$^{-1}$ exists in an area brighter than 0.9 K degrees in the velocity-declination plane. 
In the western part of N4, most CO emissions exist at 48 to 50 km s$^{-1}$. 
However, the velocity in the northwestern part of N4 changes up to 52.5 km s$^{-1}$. 
The component covering up to 52.5 km s$^{-1}$ in figure 3(b) is visible as shifting the central velocity of the spectrum to that of the red-shift side, which can be confirmed in figure 4.  
Figure 4 shows the profile map of $^{12}$CO($J$=1--0, 3--2) emission smoothed at 1 arc-min. 
Most of the $^{12}$CO($J$=1--0, 3--2) spectra are asymmetric. 
Figure 4(c) shows the spectrum of the H$_{\rm I}$ taken with the Arecibo 305 m telescope whose angular resolution is 4 arc-min, covering 4 by 4 pixels shown by the bold solid box in figure 4(b) \citep{pee18}. 
The CO emission detected here is shifted by 5 km s$^{-1}$ to the blue shift side from the peak of the H$_{\rm I}$ at 55 km s$^{-1}$, which is the nearest peak of the H$_{\rm I}$ from the velocity of CO emission.

\subsection{Intensity ratio}
Figure 5(a) shows the intensity ratio map in $^{12}$CO($J$=3--2)/$^{12}$CO($J$=1--0). 
The integrated velocity range in $^{12}$CO($J$=3--2) and $^{12}$CO($J$=1--0) is 45 to 55 km s$^{-1}$.  
The intensity ratio of $^{12}$CO($J$=3--2)/$^{12}$CO($J$=1--0) at the northern part of N4 is higher than 0.8, and also higher than that at its central and southern parts.
Figure 5(b) shows the histogram of the ratios shown in figure 5(a).
The peak of the ratio is 0.7--0.8. 
The area with the ratios higher than 1.0 dominates 5.8\% of the mapped region, while that less than 0.5 dominates 9.8\% of the region. 
Figure 5(c) shows the intensity ratio map of $^{13}$CO($J$=3--2)/$^{12}$CO($J$=3--2). 
The integrated velocity range in $^{13}$CO($J$=3--2) is also the same as the case of $^{12}$CO($J$=3--2, 1--0).
The intensity ratio of the $^{13}$CO($J$=3--2)/$^{12}$CO($J$=3--2) is higher than 0.25 at the western part of the $^{13}$CO($J$=3--2) emitting region. 
Figure 5(d) shows the histogram of the ratios shown in figure 5(c).
The area with the intensity ratio less than 0.2 distributes in the $^{13}$CO emitting region, while that with the ratio higher than 0.3 dominates 7.1\% of the region. 
The intensity ratio of the $^{13}$CO($J$=3--2)/$^{12}$CO($J$=3--2) toward N4 equals the typical value of the molecular clouds in the Galactic plane (e.g., \cite{rig16}).

\subsection{OH emission}
Figure 6 shows spectra of 6(a) $^{12}$CO($J$=1--0), 6(b) the H$_{\rm I}$ \citep{pee18} and 6(c) to 6(f) OH 1612, 1665, 1667 and 1720 MHz. 
For figures 6(a) and 6(b), the data are smoothed to the beam size of the Usuda 64m telescope for comparison. 
The H$_{\rm I}$ profile was fitted by Regularized Optimization for Hyper-Spectral Analysis (ROHSA) \citep{mar19}. 
The gaussian fit was conducted to the wide region, including the position of the current OH observation, and only the result of the fit toward the position of the OH observation is shown in the figure. 
The H$_{\rm I}$ profile is divided into several components, and the sum of the divided spectra reproduces the observed spectrum. 
Any OH emission at 3 bands, 1612 MHz, 1665 MHz, and 1720 MHz was not detected toward N4 over the 3-sgima r.m.s. noise level of $\sim$0.3 K. 
Emission at 47 km s$^{-1}$ and absorption at $\sim$52 and $\sim$53 km s$^{-1}$ are shown at only 1667 MHz. 
The details of the emission and absorption are discussed in section 4.4.

\section{Discussion}

\subsection{Physical properties of the molecular cloud}

\subsubsection{RADEX analysis of the molecular cloud}
Multi transition and isotope of the CO emission can trace the excitation condition of the CO in the molecular cloud. 
In order to constrain the physical properties of the molecular cloud through the intensity ratios, here we apply the RADEX code, which is the non-local thermodynamic equilibrium radiative transfer code \citep{van07}. 
This code calculates the line intensity for an input set of a kinetic temperature ($T_{\rm k}$), a volume number density of H$_2$ ($n$(H$_2$)), a column density, and a line width of the molecule whose line intensity is calculated. 
Liden Atomic and Molecular Database were used to input collision rates of the molecules with hydrogen molecules \citep{sch05}. 
For the input parameter sets of $T_{\rm k}$ and $n$(H$_2$) changing at a range of 10 K to 10$^2$ K every 10$^{0.025}$ K and 10$^2$ cm$^{-3}$ to 10$^6$ cm$^{-3}$ every 10$^{0.05}$ cm$^{-3}$, the intensities of $^{12}$CO($J$=1--0), $^{12}$CO($J$=3--2), and $^{13}$CO($J$=3--2) emissions are calculated. 
Then, line ratios of $^{12}$CO($J$=3--2)/$^{12}$CO($J$=1--0), $^{13}$CO($J$=3--2)/$^{12}$CO($J$=3--2), and $^{13}$CO($J$=3--2)/$^{12}$CO($J$=1--0) were calculated in the $n$(H$_2$)$-$$T_{\rm k}$ space. 
The parameters of the abundance ratio of $^{12}$CO/$^{13}$CO = 60, $N$($^{12}$CO) = 5$\times$10$^{17}$ cm$^{-2}$ are applied. 
Eight positions shown in the upper-left in figure 7 are selected to calculate the $T_{\rm k}$ and $n$(H$_2$) of the molecular cloud by the RADEX code. 
These points are included in the local peak in $^{12}$CO($J$=1--0) emission or $^{13}$CO($J$=3--2) emission. 
The observed line intensity ratio in each of the eight positions, which is shown in table 1, with 10\% error, is plotted in $n$(H$_2$)$-$$T_{\rm k}$ space in each panel (A) to (H) in figure 7. 
The Chi-square test is also carried out simultaneously, and the point of minimum value of the Chi-square in $n$(H$_2$)$-$$T_{\rm k}$ space is adopted as the solution of the RADEX calculation. 
Table 1 shows the results of the RADEX calculation in the eight positions. 

We find that $T_{\rm k}$ and $n$(H$_2$) that can explain the observed line intensity ratio ranges from $\sim$15 K to $\sim$56 K and from $\sim$1.4$\times$10$^3$ cm$^{-3}$ to $\sim$1.8$\times$10$^4$ cm$^{-3}$, respectively. 
The western part of the cloud, B, D, E, and G, whose $T_{\rm k}$ and $n$(H$_2$) range from $\sim$24 K to $\sim$56 K and from $\sim$2.5$\times$10$^3$ cm$^{-3}$ to $\sim$1.8$\times$10$^4$ cm$^{-3}$, respectively, seems to have denser gas and higher $T_{\rm k}$ than those of the eastern part of the cloud, A, C, F, and H whose $T_{\rm k}$ and $n$(H$_2$) range $\sim$15 K to $\sim$24 K and $\sim$1.4$\times$10$^3$ cm$^{-3}$ to $\sim$2.8$\times$10$^3$ cm$^{-3}$. 
As a whole, $T_{\rm k}$ of N4 is higher than that of typical molecular clouds in the Galactic Plane.
The western part of N4 has an exceptionally high $T_{k}$ up to $\sim$56 K, meaning that N4 is heated up.
The details of the heating mechanism are discussed in section 4.2.

\subsubsection{Mass estimation}
The mass of molecular cloud, $M_{\rm mol}$, is estimated by
\begin{equation}
N({\rm H_2}) = W({\rm ^{12}CO}(J=1-0)) \times  X_{\rm CO}
\end{equation}
\begin{equation}
M_{\rm mol} = \overline{\mu} m_{\rm H} \Sigma[N({\rm H_2}) \times  \Omega \times D^2],
\end{equation}
where $X_{\rm CO}$ is the conversion factor of $N$(H$_2$) from the intensity of CO emission, $\overline{\mu}$ the mean molecular weight, $m_{\rm H}$ the mass of the atomic hydrogen, $\Omega$ the solid angle subtended by the grid spacing, and $D$ the distance to the molecular cloud. 
Here, we adopt that $X_{\rm CO}$ = 2$\times$10$^{20}$ cm$^{-2}$/(K km s$^{-1}$) \citep{bol13}, $\overline{\mu}$ = 2.8 when taking into account a helium abundance of 25\% of mass, and $D$ = 5500 pc. 
The summation is conducted over the observed area where the integrated intensity is higher than 3$\sigma$ (5.7 K km s$^{-1}$) in $^{12}$CO($J$=1--0). 
Then, the mass of the molecular cloud is estimated to be $\sim$7300 $M_\odot$. 

Here we verify the mass of the molecular cloud. 
Given LTE (local thermodynamic equilibrium) in the molecular cloud, an excitation temperature, an optical depth of the CO emission, and a column density of the CO molecule, $N$(CO) are calculated by

\begin{equation}
T_{\rm ex} = \frac{h\nu_{\rm CO}}{k} \left[{\rm ln}\left\{ \frac{1}{\frac{kT_{\rm MB}}{h\nu_{\rm CO}}\frac{1}{1-{\rm exp}\left(-\tau_{\rm CO}\right)} + \frac{1}{{\rm exp}\left(\frac{h\nu_{\rm CO}}{kT_{\rm bg}}\right) -1}} +1 \right\}  \right]^{-1}
\end{equation}

\begin{equation}
\tau_{\rm CO} = -{\rm ln}\left[1-\frac{kT_{\rm MB}}{h\nu_{\rm CO}}\left\{\frac{1}{{\rm exp}\left(\frac{h\nu_{\rm CO}}{kT_{\rm ex}}\right)-1} - \frac{1}{{\rm exp}\left(\frac{h\nu_{\rm CO}}{kT_{\rm bg}}\right)-1}\right\}^{-1}\right]
\end{equation}
\begin{equation}
N({\rm CO}) = \frac{3ckT_{\rm ex}{\rm d} V \tau_{\rm CO}}{8\pi^3 B \mu^2\left\{1-{\rm exp}\left(-\frac{h\nu_0}{kT_{\rm ex}}\right)\right\}} \frac{2J+1}{J+1} \frac{2J+1}{2J+3} \frac{1}{{\rm exp}\left\{-\frac{h B J(J+1)}{kT_{\rm ex}}\right\}}
\end{equation}
where, $h$ is the Planck constant, $\nu_{\rm CO}$ the frequency of the CO emission, $k$ the Boltzmann constant, $\tau_{\rm CO}$ the optical depth of the CO emission, $T_{\rm bg}$ = 2.7 K the temperature of the cosmic microwave background, $c$ the light speed, $B$ the rotational constant of the CO emission and $\mu$ the electric dipole moment.
In the cloud, $T_{\rm ex}$ is estimated to be $\sim$33 K by assuming that the $^{12}$CO emission is optically thick and $T_{\rm mb}$ in $^{13}${\rm CO}($J$=3--2) is $\sim$6 K. 
Then, $\tau_{^{13}CO(J=3-2)}$ is estimated to be $\sim$0.27 when assuming the same $T_{\rm ex}$ between $^{12}$CO and $^{13}$CO. 
When a line width of the $^{13}$CO($J$=3--2) emission is assumed to be $\sim$0.5 km s$^{-1}$, $N$($^{13}$CO) is estimated to be $\sim$1.63$\times$10$^{16}$ cm$^{-2}$. 
By applying the abundance ratio of H$_2$/$^{13}$CO of 7$\times$10$^5$ \citep{dic78}, $N$(H$_2$) is estimated to be $\sim$1.1$\times$10$^{22}$ cm$^{-2}$. 
In addition, toward the same position, the integrated intensity of $^{12}$CO($J$=1--0) emission is estimated to be $\sim$50 K km s$^{-1}$. Then, $N$(H$_2$) is estimated to be $\sim$1.0$\times$10$^{22}$ cm$^{-2}$ using $X_{\rm CO}$ of 2$\times$10$^{20}$ cm$^{-2}$/(K km s$^{-1}$). 
When the RADEX result does not assume the LTE, $N$(H$_2$) is estimated to be $\sim$1.2$\times$10$^{22}$ cm$^{-2}$. 
The other positions also show the same tendency. 
These results may indicate that the $N$(H$_2$) of 1--2$\times$10$^{22}$ cm$^{-2}$ is plausible toward this position, and the mass of 7300 $M_\odot$ estimated from this $N$(H$_2$) value is also plausible. 

\subsubsection{Internal motion}
In figure 3, the main body of N4 has a velocity gradient of  $\sim$0.16 km s$^{-1}$ (20$\arcsec$)$^{-1}$, and the northern part of N4 is a more red-shifted. 
Furthermore, the northwestern part of N4 has more red-shifted velocity components up to 3 km s$^{-1}$ from the systemic velocity. 
In almost all region of N4, CO spectra show the asymmetric shape for the LSR velocity (See figure 4), meaning that the non-thermal shock affects N4. 
The location of N4 in figure 1 is along the same line of sight as radio shell W50. 
N4 has the same velocity as N1, N2, and N3, and a detailed analysis of N1, N2, and N3 by \citet{liu20} indicates that N1, N2, and N3 are associated with the SS433/W50 system.
These results indicate that N4 is also associated with the SS433/W50 system.

The thermal pressure calculated from RADEX results is listed in table 1. 
The thermal pressure at the western part of the molecular cloud, B, D, E, and G, which is at an order of 10$^5$ K cm$^{-3}$, is also higher than that at the eastern part. 
Turbulent pressure calculated by the formula (14) in \citet{rig19} for N4 is estimated to be $\sim$9.0$\times$10$^6$ K cm$^{-3}$, where the mean line width of $^{12}$CO($J$=1--0) emission  is 1.8 km s$^{-1}$, excitation temperature is $\sim$30 K and average $n$(H$_2$) $\sim$2700 cm$^{-3}$ estimated by the RADEX results excluding G due to large error.  
Figure 8(a) shows the plots of turbulent and thermal pressure. 
Black plots in the figure are for the molecular clouds at 1st quadrant derived by CHIMPS \citep{rig19}, and the red plot is the case of N4. 
As a general trend, the turbulent pressure is one to two orders of magnitude higher than the thermal pressure in almost all molecular clouds. 
N4 also has the same trend and is located in the high turbulent pressure area in the plot. 
Figure 8(b) shows the plots of turbulent pressure and mass of the molecular clouds. 
The black and red plots are the same in figure 8(a).  
The turbulent pressure of the molecular clouds ranges from 10$^{4}$ to 10$^{7}$ K cm$^{-3}$ and does not depend on their mass. 
N4 is plotted at the high turbulent pressure area compared with the molecular clouds of the same mass by CHIMPS. 
These results indicate that the thermal and turbulent pressures in N4 have the same trend as the molecular clouds in the Galactic Plane, and those are higher than the molecular clouds. 
This is due to the strong shock from the SS433/W50 system, supporting the association of the SS433/W50 system.   

\subsubsection{Strength and distribution of CO emission}

Figure 9 shows the distribution of peak main-beam temperature, $T_{\rm MB}$, in 9(a) $^{12}$CO($J$=1--0) and 9(b) $^{12}$CO($J$=3--2) emissions. 
The brightest position of $T_{\rm MB}$ in $^{12}$CO($J$=1--0) emission is located at $\sim$1 pc from the northern edge of N4, indicating that a  large $T_{\rm MB}$ gradient exists at the northern edge at a 3$\sigma$ level. 
The peak $T_{\rm MB}$ in $^{12}$CO($J$=1--0) is high at the western part of the molecular cloud compared with the eastern part, and it exceeds 30 K at the western part. 
The region where $T_{\rm MB}$ greater than 25 K is distributed at 2 pc $\times$ 2 pc at the western part.
The distribution of the peak $T_{\rm MB}$ in $^{12}$CO($J$=3--2) emission tends to be similar to that in $^{12}$CO($J$=1--0) emission, and peak $T_{\rm MB}$ in $^{12}$CO($J$=3--2) emission also exceeds 30 K at the peak. 
A large $T_{\rm MB}$ gradient is also seen at the northern edge of the molecular cloud, and the brightest position of $T_{\rm MB}$ in $^{12}$CO($J$=3--2) emission is located within 1 pc of the edge of N4. 
A $T_{\rm MB}$ gradient is also seen at the eastern and western edge of N4 as well as the northern one. 
The peak $T_{\rm MB}$ distribution with a large gradient can be seen only at the molecular clouds associated with the H$_{\rm II}$ region or supernova remnants where strong interaction with the ISM occurs (e.g., \cite{tac00}). 
The integrated intensity of the $^{12}$CO($J$=1--0, 3--2) emission also has a strong intensity gradient from the edge at a 3$\sigma$ level. 
These indicate that the envelope of N4 has been stripped by the strong shock of a supernova explosion and/or other shock events passing through N4, and only a dense part of the cloud survives.
The same case of the large CO intensity gradient was reported in the SNR RX J1713.7-3946 \citep{san10}. 
They concluded that a molecular cloud exists within or at the outer boundary of the SNR shell. 
This conclusion may be applicable to the case of N4.

\subsection{Shock interaction of a cloud with the jet from SS433}
In section 4.1, the association of N4 with the SS433/W50 system is briefly discussed. 
In this subsection, we further discuss the association and the origin of the shock. 
Our RADEX analysis suggests that $T_{\rm k}$ in N4 is higher than that of typical molecular clouds as a whole and, particularly in the western part of the molecular cloud, it is as high as $T_{\rm k}$ of $\sim$50 K. 
The age of the W50 is estimated to be $\sim$10$^5$ yr (e.g., \cite{pan17}).
The cooling timescale of the dense cloud with a density of $\sim$10$^3$ cm$^{-3}$, is $\sim$10$^3$ yr (e.g., \cite{koy00}). 
The projected distance to N4 from SS433 is $\sim$40 pc. 
Given the average shock velocity of the supernova explosion of W50 of $\sim$1000 km s$^{-1}$, a travel time of 4.3$\times$10$^4$ yr is required for the shock of the supernova to reach N4. 
Considering these timescales, if the input energy to N4 is provided by only a supernova shock, then N4 should have been already cooled down. 
In order to explain the high $T_{\rm k}$ in N4 as high as $\sim$56 K, an external force should be still working at present. 
Figure 10 shows the distribution of the mid/far-infrared radiation (a: WISE 12$\mu$m, b: WISE 22$\mu$m, c: AKARI WIDE-S, d: AKARI WIDE-L) superposed on the peak $T_{\rm MB}$ distribution of $^{12}$CO($J$=1--0) emission in contours. 
Diffuse mid/far-infrared radiation surrounds N4, and a strong one is only seen at the northwestern part of N4 where the $T_{\rm MB}$ is high.
The distribution of the strong mid/far-infrared radiation does not coincide with that of $T_{\rm MB}$ in $^{12}$CO. 
However, the edge of the region where the strong mid/far infrared radiation is detected coincides with the edge of the CO emitting region at the northern and western part of N4, nearly corresponding to the position where a peculiar velocity shifted by up to 3 km s$^{-1}$ appears (figure 3(b)). 
The WISE 12$\mu$m emission races the PAH emission, which is formed where the interstellar dust is destroyed. 
The WISE 22$\mu$m, the AKARI WIDE-S, and the AKARI WIDE-L trace the thermal radiation from interstellar dust grains. 
These indicate that the mid/far-infrared radiation is associated with N4.
Figure 11 shows an image of X-ray radiation taken by ROSAT and superposed on the distribution of $^{12}$CO integrated intensity in contours.
N4 is located toward the same line of sight as that of X-ray radiation, which traces the jet from SS433.

Here we summarize the circumstantial evidence for shock interaction in N4 as follows,
\begin{enumerate}
\item Asymmetric spectra in CO emission over the whole molecular cloud
\item A velocity gradient for the systemic velocity of N4 and its shift by up to 3 km s$^{-1}$ at only the western part of N4
\item High peak $T_{\rm MB}$ and $T_{\rm k}$, not cooling down in spite of taking longer timescale than the cooling time scale
\item Higher turbulent and thermal pressures than those of the molecular clouds in the Galactic Plane
\item No diffuse envelope in N4 and located within the radio shell or at the boundary of the radio shell
\item The association with the strong mid/far-infrared radiation, which traces the PAH emission and thermal radiation at high $T_{\rm MB}$, and the region at a velocity shifted by up to 3 km s$^{-1}$
\item Same line of sight as that of the X-ray jet  
\end{enumerate}
Several pieces of circumstantial evidence strongly suggest that N4 is interacting with the jet from SS433.
Star formation is also considered another case to explain the circumstantial evidence for N4. 
However, there are no point sources found by GAIA, 2MASS, IRAS, and AKARI survey. 
Therefore, the possibility of star formation is ruled out. 
Cosmic-ray heating also has the potential to keep high $T_{\rm k}$ in N4. 
\citet{liu20} estimated the energy consumed to keep the high temperature by cosmic-ray heating at N3 in SS433.
They concluded that cosmic-ray heating can keep high temperature in the cloud because this energy at N3 was estimated to be an order of 10$^{46}$ ergs, which is a few orders of magnitude lower than the canonical energy for the cosmic rays accelerated by the SNR shock.
In the case of N4, if applying the same method as \citet{liu20} to estimate the energy to keep high $T_{\rm k}$ by cosmic-ray heating, the energy was estimated to be an order of 10$^{47}$  ergs.
Therefore, contribution of cosmic-ray heating to keep $T_{\rm k}$ $\sim$50 K can not be ruled out.
By considering the situation of N4 and the jet from SS433, the strong interaction may have continued for more than several thousand years. 
N4 would be located in front of the X-ray jet, which is expected from the velocity structure of N4.
This result is also based on the detailed comparion of N4 with X-ray image by another paper (in preparation).
The energy of the component whose systemic velocity is shifted at the northwestern part of N4 is estimated to be $\sim$3$\times$10$^{47}$ ergs. 
This can easily explain the cause of the velocity shift because the kinetic energy of the jet has been estimated to be 3.2$\times$10$^{38}$ ergs s$^{-1}$ \citep{mar02}, corresponding to 1.0$\times$10$^{46}$ ergs yr$^{-1}$. 
However, we need numerical simulation to accurately understand why the velocity shift is $\sim$3 km s$^{-1}$ by an interaction with the jet.

Figure 12 shows the schematic view of the interaction of N4 with the jet from SS433. 
N4 is located just in front of the jet and interacts with the jet from backward of the line of sight. 
The timescale of the jet precession was measured to be $\sim$162.5 d \citep{blu04}. 
If the jet keeps the velocity of 0.26$c$ at a distance of $\sim$40 pc from SS433, the jet is propagated to a distance of only $\sim$0.037 pc at a single precession cycle, corresponding to $\sim$1.4 arc-sec at 5.5 kpc. 
This means that in pc-scale view, the shape of the jet is a hollow cylinder shown in figure 12(a). 
As represented in figure 12(b), most CO spectra are asymmetric. 
Wing components at the blue-shifted velocity side are shown at the eastern part of N4, and those at the red-shifted velocity side are seen at only the western part of N4. 
Between them, wing components at both velocity sides can be seen, but only in small areas. 
At the most western part of N4, the centroid velocity of the spectra is shifted to the red-shifted side, which is also seen in the position-velocity diagram in figure 3.  
Figure 12(c), 12(d), and 12(e) show the schematic view of the interaction seen from the direction 12(c): the front, 12(d) top, and 12(e) west side, respectively. 
The blue and red arrows indicate the directions of the internal motions of N4 to explain the CO spectra in figure 12(b). 
The eastern part of N4 is moving to us and being pushed by the jet. 
On the other hand, toward the northwestern part of N4, if we explain velocity shifted to red-shift side in current condition, it is natural that red-shifted gas is moved by rotating motion of the jet formed by the precession of compact star toward cross-section for propagating direction of the jet.
This motion of N4 couples the magnetic field.
The strong interaction by engulfment with the magnetic field would cause the high temperature of the gas. 
This result may help further discussion of the nature of the jet from SS433.

\subsection{Comparison with high energy radiation}
The dense ISM affecting the strong shock is possibly at the site of the cosmic ray particle acceleration \citep{fer49}. 
This can often be seen toward the supernova remnant (e.g., \cite{san21}). 
There gamma-ray radiation up to TeV was detected, and the origin of the TeV gamma-ray was discussed to result from the interaction of high energy cosmic-ray proton with an interstellar proton. 
Various gamma-ray observations have been conducted toward the SS433/W50 system (e.g., \cite{abe18}; \cite{fan20}). 
Actually, toward the jet from SS433, Gev-TeV gamma-ray have been detected, and their origins have been discussed as the leptonic induced by inverse Compton \citep{abe18}. 
However, any GeV-TeV gamma-ray radiation has not been detected toward N4. 
In the case of W44, the TeV gamma-ray radiation by the hadronic process is detected at the edge of the shell where the strong 1.4 GHz radio continuum was also detected \citep{yos13}. 
The 1.4 GHz radio continuum is radiated by the synchrotron radiation, meaning that there exists relatively a strong magnetic field. 
The particle acceleration needs strong magnetic field in different two-way fluids. 
In figure 1, 1.4 GHz radio continuum is not or very weakly detected toward N4. 
This may indicate the magnetic field is weak or the radio continuum emission tracing the magnetic field already disappeared by synchrotron radiation. 
The GeV gamma-ray emission has recently been detected toward the jet from SS433 (\cite{abe18}; \cite{fan20}; \cite{li020}). 
N4 is located at the edge of the distribution of the gamma-ray emission. 
The observing area by the NRO45m telescope covers the peak direction of the gamma-ray emission. 
However, no CO emission could be detected, suggesting that there are no dense molecular clouds there. 
As shown in figure 4(c), the H$_{\rm I}$ spectrum is complicated, making it difficult to discuss the warm and the cold neutral medium of the H$_{\rm I}$ at the gamma-ray radiating direction. 
\citet{liu20} discussed the possibility to detect the gamma-ray radiation by the hadronic process toward N2 and N3. 
Their calculation suggests that, even if the gamma-ray by the hadronic process is radiated, it is too weak to detect it by the current gamma-ray observatory; it is challenging even with CTA. 
Considering such a situation, it may also be difficult to detect the gamma-ray radiation by the hadronic process toward N4. 
However, further study concerning the dense cloud and diffusive shock acceleration is needed.

\subsection{Non-detection of the OH emission}
No apparent emission and absorption was detected at 1612 MHz, 1665 MHz, and 1720 MHz toward N4. 
According to the result of the Gaussian fit of the H$_{\rm I}$ spectrum in figure 6(b), no H$_{\rm I}$ component exists at $\sim$47 km s$^{-1}$ where the emission-like feature is seen at 1667 MHz. 
The velocity of $\sim$47 km s$^{-1}$ is also different from that of the emission in $^{12}$CO($J$=1--0). 
The Velocities of $\sim$52.2 km s$^{-1}$ and 53.3 km s$^{-1}$, in which the absorption-like features at 1667 MHz are seen, corresponds to those of H$_{\rm I}$ components derived by the Gaussian fitting with ROHSA \citep{mar19}. 
However, the line width of the absorptions is less than 0.1 km s$^{-1}$, too narrow to consider that the absorptions are astronomical signals. 
Therefore, these emission and absorption at 1667 MHz are considered to be artificial signals, and as a result, no OH emission is confirmed toward N4. 

There are two cases to emit the OH emission there, one is maser emission by the strong shock between the jet, and dense molecular clouds and other is thermal OH emission in the ISM, including the dense molecular cloud.
OH maser emission at 1720 MHz is a good tracer of C-shock \citep{loc99}. 
This maser emission is often seen in the SNRs (e.g., \cite{bro00}). 
The OH maser at 1720 MHz is emitted in the region where a gas temperature,$T$, ranges 50 to 125 K, volume density of hydrogen molecules, $n$(H$_2$), is an order of 10$^5$ cm$^{-3}$, and a collumn density of OH molecules, $N$(OH), is an order of 10$^{16}$ cm$^{-2}$ \citep{loc99}. 
The upper limit of $N$(OH) in N4 is estimated to be $\sim$1.6$\times$10$^{15}$ cm$^{-2}$. 
$n$(H$_2$) and $N$(OH) are lower than the ranges necessary to emit the OH maser at 1720 MHz. 
Therefore, it can be explained why OH maser at 1720 MHz could not be detected in the present observation. 

\citet{rug18} compared $N$(OH) with $N$(H$_2$) in the Galactic plane as a part of the THOR project (\cite{beu16}; \cite{wan20}). 
They found the relationship of 
\begin{equation}
{\rm log}(N({\rm OH})) = 0.33^{+0.14}_{-0.13} \times {\rm log}(N({\rm H_2})) + 7.91^{+2.86}_{-2.95}
\end{equation}
by combining equations (1) and (6), then
\begin{equation}
{\rm log}(N({\rm OH})) = 0.33^{+0.14}_{-0.13} \times {\rm log}(W_{\rm CO}) + 14.61^{+2.86}_{-2.95}
\end{equation}
If it here assumes that the OH emission is only radiated within the CO cloud, from equation (7), the smoothed $N$(OH) at Usuda 64m beam in N4 is estimated to be $\sim$3.5$\times$10$^{14}$ cm$^{-2}$. 
The equation by estimating $N$(OH) is also given by
\begin{equation}
N({\rm OH}) = 2.39\times10^{14} \times T_{\rm ex}(1667) \times \tau(1667) \times \Delta V,
\end{equation}
where $T_{\rm ex}$ is the excitation temperature of the OH molecule, $\tau$ is an optical depth at the peak of the Gaussian profile, and $\Delta V$ is the line width of the OH emission deriving by Gaussian fit. 
The typical derived $\tau$(1667) of most of the clouds in the Galactic plane listed in table 2 in \citet{rug18} is less than 1, 0.21 on average and 0.12 in median value. 
Therefore, here giving the $\tau$(1667) of 0.12 and $\Delta V$ of $\sim$1.8 km s$^{-1}$, equal to an average line width of CO emission under assuming that line width of OH and CO is similar in a turbulent dominated gas, then $T_{\rm ex}$ is calculated to be $\sim$7.2 K. 
The observed brightness temperature, $T_{\rm b}$ is formulated using the following equation,
\begin{equation}
T_{\rm b} = (T_{\rm ex} - T_{\rm bg}) \times (1 - e^{-\tau}).
\end{equation}
Here the background temperature, $T_{\rm bg}$, is adopted from the estimate by the CHIPASS project taken with Parkes 64 m telescope \citep{cal14}. 
Although the observed frequency of the CHIPASS was 1.4 GHz, $T_{\rm bg}$ at 1.6 GHz is assumed to be equal to that at 1.4 GHz. 
The $T_{\rm bg}$ toward the observing point by Usuda 64 m telescope was $\sim$8.9 K. 
Then, $T_{\rm b}$ is estimated to be $\sim$190 mK. 
Taking into account the r.m.s. noise level of the data in OH emission and the main beam efficiency of the Usuda 64 m telescope, this level of emission can not be detected by the current short integration observations. 
Therefore, we need to update the observation system and more integration time to detect fainter OH emissions.  

In order to reveal the structure of low-density gas at the resolution similar to that of CO, $\sim$15$\arcsec$, the Square Kilometre Array (SKA) is the best system to be observe it. 
However, even at the sensitivity of the SKA at $\sim$1.6 GHz band, detection of thermal OH emission as strong as a few tens of mK is still challenging at $\sim$0.1 km s$^{-1}$ resolution without spatially smoothing.  

\section{Summary}
We conducted observations and analyses of the molecular cloud, N4 at the western part of SS433 in $^{12}$CO($J$=1--0, 3--2) and $^{13}$CO($J$=3--2) and four lines of the grand-state of OH emission. 
The summary of the paper is described as follows;

\begin{enumerate}
   \item Three out of four molecular clouds have a good spatial correlation with the radio shell in 1420MHz. One molecular cloud, N4 is located at the same line of sight as the radio shell.
   \item The integrated intensity distribution of $^{12}$CO($J$=1--0, 3--2) emission in N4 has two local peaks. A peak of $T_{\rm MB}$ is seen at the western part of N4, while a strong $T_{\rm MB}$ gradient is seen at the edge of N4 at the northern, eastern, and western parts.
   \item A velocity gradient of $\sim$0.16 km s$^{-1}$ (20$\arcsec$)$^{-1}$ is seen in N4. At the northwestern part of N4, a peculiar velocity shift by up to 3 km s$^{-1}$ is seen. 
   \item Using the RADEX code, The $n_{\rm (H_2)}$ and $T_{\rm k}$ are estimated at 8 positions toward  the local peak of $^{12}$CO($J$=1--0) and $^{13}$CO($J$=3--2). The calculated $n_{\rm (H_2)}$ is an order of 10$^3$ cm$^{-3}$, and $T_{\rm k}$ ranges $\sim$20 to $\sim$56 K, and $T_{\rm k}$ is high at the western part of N4.
   \item The mass of the molecular clouds is estimated to be $\sim$7300 $M_{\rm \odot}$.
   \item The thermal and turbulent pressures in N4 are estimated to be $\sim$10$^5$ K cm$^{-3}$ and 10$^7$ K cm$^{-3}$, respectively. The relation of thermal and turbulent pressures tends to be similar to that of the molecular clouds in the Galactic plane, but the turbulent and thermal pressures of N4 is higher than the typical molecular clouds in the Galactic plane. 
   \item The following pieces of circumstantial evidence for N4 suggest that N4 is interacting with the jet from SS433.
   \begin{itemize}
      \item Asymmetric spectra in CO emission over the whole molecular cloud
      \item A gradient of the systemic velocity of N4 and a velocity shifted by up to 3 km s$^{-1}$ at only the western part of N4
      \item High peak $T_{\rm MB}$ and $T_{\rm k}$ indicating non-cooling down in spite of longer timescale of W50 than cooling time scale
      \item Higher turbulent and thermal pressures than those in the molecular clouds in the Galactic Plane
      \item No diffuse envelope around N4 and within and over the boundary of the radio shell
      \item Association of CO emission with the strong mid/far-infrared radiation which traces the PAH emission and thermal radiation at high $T_{\rm MB}$, and the region with a velocity shifted by up to 3 km s$^{-1}$
      \item CO emission in the same line of sight as that of the X-ray jet  
   \end{itemize}
   \item No-gamma ray radiation is detected toward N4. From comparing the possibility to detect the gamma-ray radiation generated by the interaction of the cosmic-ray protons from N2 and N3 with ISM protons in from N4, it is concluded that it is difficult to detect the gamma-ray radiation toward N4.
   \item OH emission at 1612, 1665, 1667, and 1720 MHz has not been detected toward N4. This is probably due to the low sensitivity of the observation and beam dilution.
\end{enumerate}


\clearpage

\begin{figure}
  \begin{center}
    \includegraphics[width=160mm]{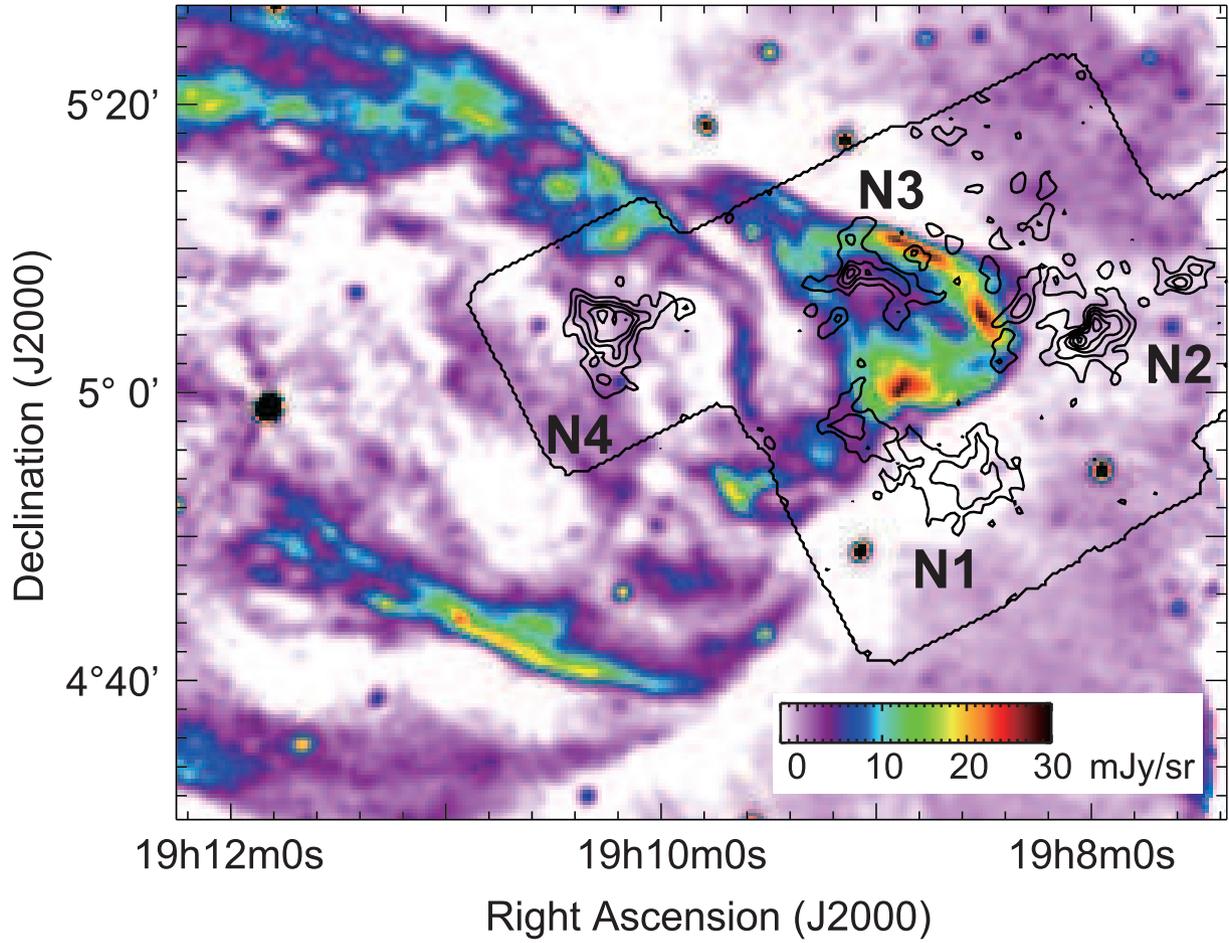}
  \end{center}
  \caption{Integrated intensity map of $^{12}$CO($J$=1--0) emission observed with the NRO45m in contours toward the western part of the SS433/W50 system superposed on the radio continuum at 1.45 GHz observed with the VLA in a color image \citep{dub98}. The integrated velocity range of $^{12}$CO emission is 45 km s$^{-1}$ to 55 km s$^{-1}$. The contour levels of the $^{12}$CO emission increase from 10 K km s$^{-1}$ to 75 K km s$^{-1}$ by a step of 13 K km s$^{-1}$. An enclosure line in figure shows the observed region with the NRO45m.  \label{fig:f1}}
\end{figure}

\begin{figure}
  \begin{center}
    \includegraphics[width=160mm]{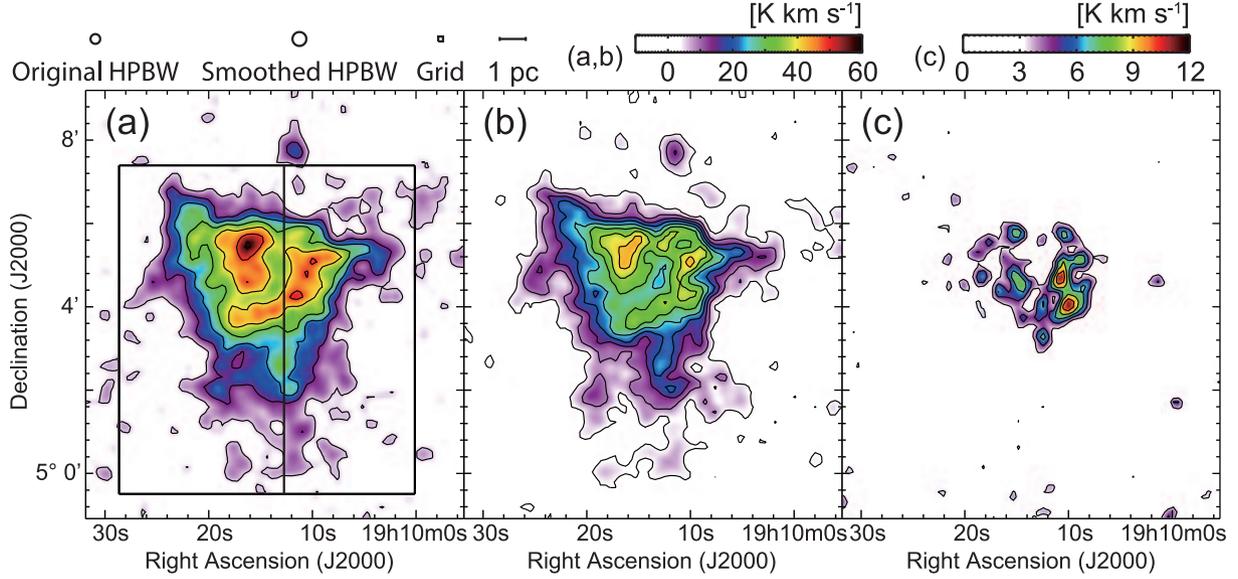}
  \end{center}
  \caption{Velocity integrated intensity maps of (a) $^{12}$CO($J$=1--0), (b) $^{12}$CO($J$=3--2) and (c) $^{13}$CO($J$=3--2) emission. Observations of $^{12}$CO($J$=1--0) are taken with the NRO45m, and those of $^{12}$CO($J$=3--2), and $^{13}$CO($J$=3--2) emission are taken with the JCMT15m. The integrated velocity range is 45 km s$^{-1}$ to 55 km s$^{-1}$ in all three maps. The contour levels in (a), (b), and (c) increase from 3$\sigma$(5.7 K km s$^{-1}$) to 30$\sigma$ by a step of 4.5$\sigma$, from 3$\sigma$(3.6 K km s$^{-1}$) to 34.5$\sigma$ by a step of 4.5$\sigma$ and from 3$\sigma$(3.3 K km s$^{-1}$) to 9$\sigma$ by a step of 1.5$\sigma$, respectively. The beam sizes of the NRO45m in CO($J$=1--0) and JCMT15m in CO($J$=3--2) are nearly equal and the grid size is also set to equal to 7.2 arcseconds between the observations with both the NRO45m and the JCMT15m. The beam sizes of all the three datasets smoothed to 20 arcseconds, the original beam sizes before smoothing, and the mapping grid size are displayed at the upper left of the figure. The color scales in (a) and (b) are equal. Two solid boxes in (a) show the area the making position-velocity maps in figure 3.  \label{fig:f2}}
\end{figure}

\begin{figure}
  \begin{center}
    \includegraphics[width=160mm]{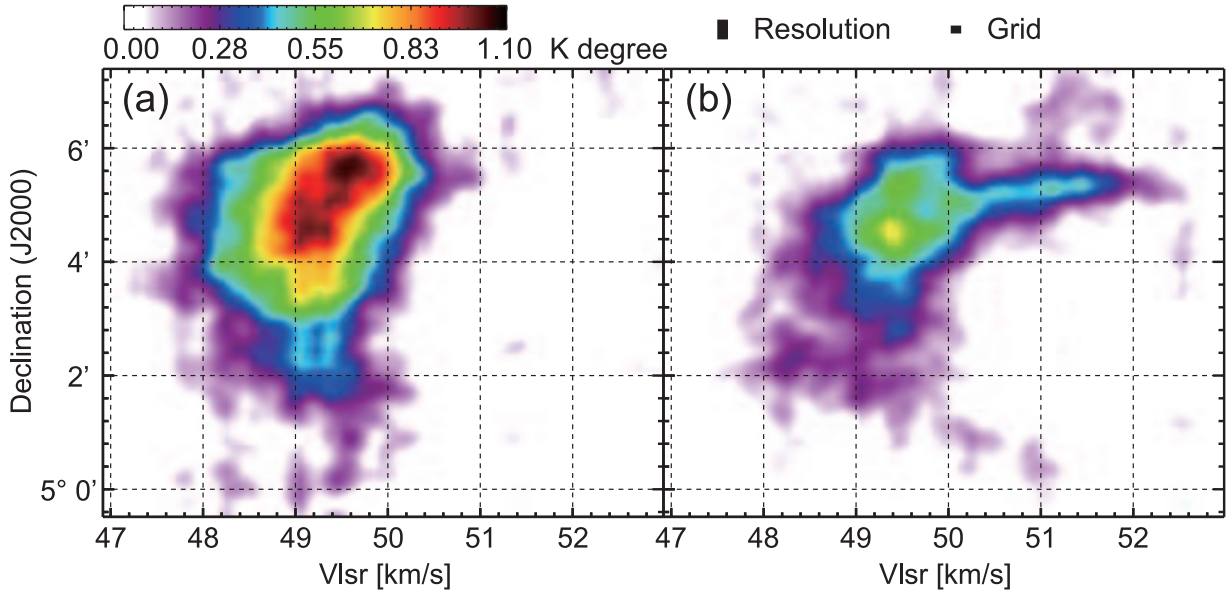}
  \end{center}
  \caption{Position-velocity diagrams of CO emission synthesized with the data in the (a) left and (b) right boxes in figure 2(a), respectively. The data is integrated along the Right Ascension.   \label{fig:f3}}
\end{figure}

\begin{figure}
  \begin{center}
    \includegraphics[width=160mm]{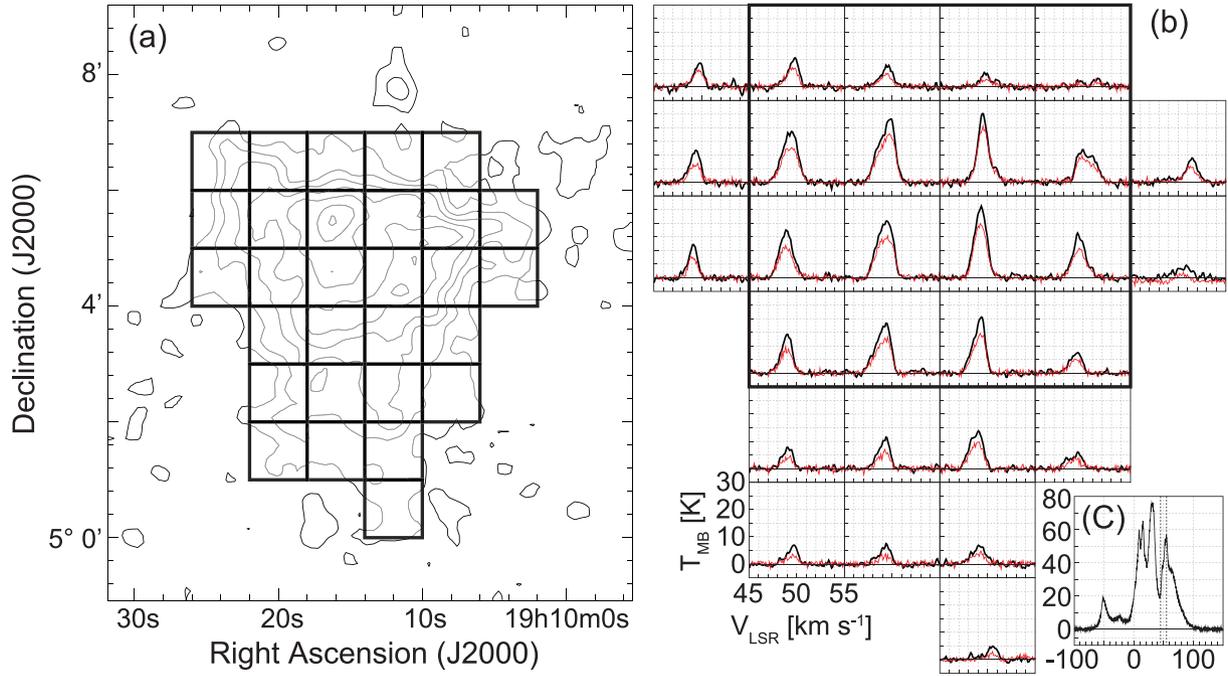}
  \end{center}
  \caption{(a) Integrated intensity map of $^{12}$CO($J$=1--0) emission superposed on the grids to show the $^{12}$CO($J$=1--0) and $^{12}$CO($J$=3--2) spectra in (b). The integrated velocity range and contour levels of $^{12}$CO($J$=1--0) emission are the same in figure 2(a). The grid size is 1 arc-min. (b) The averaged profile maps in $^{12}$CO($J$=1--0) (black) and $^{12}$CO($J$=3--2) (red) in each grid of 1 square arc-min. The bold solid box by 4 $\times$ 4 pixels is the area of the profile of the H$_{\rm I}$ shown in (c), whose spatial resolution is 4 arc-min with the Arecibo 305 m telescope.  (c) The profile map of the H$_{\rm I}$ toward the bold solid box in (b). The vertical bold dotted lines show the min/max range of the velocity of the profile maps in (b).    \label{fig:f4}}
\end{figure}

\begin{figure}
  \begin{center}
    \includegraphics[width=160mm]{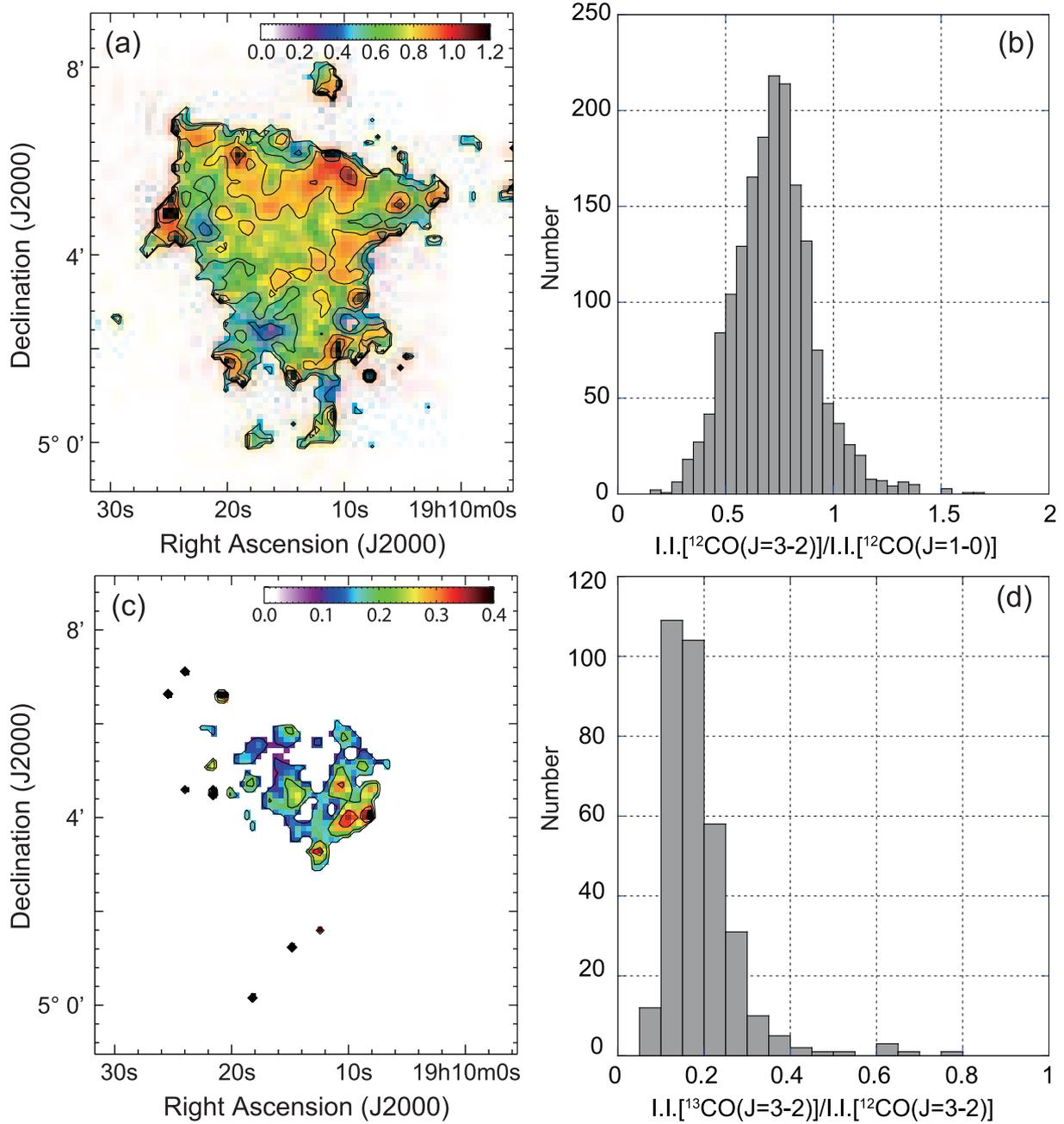}
  \end{center}
  \caption{Maps of the intensity ratio and histgrams. The integrated velocity range in $^{12}$CO($J$=3--2) and $^{12}$CO($J$=1--0) used here is 45 to 55 km s$^{-1}$.  (a), (b) Those for $^{12}$CO($J$=3--2) and $^{12}$CO($J$=1--0). The contour levels in (a) increase from 0.4 to 1.0 by a step of 0.2.  (c), (d) Those for $^{13}$CO($J$=3--2) and $^{12}$CO($J$=3--2). The contour levels in (c) increase from 0.1 to 0.4 by a step of 0.1.    \label{fig:f5}}
\end{figure}

\begin{figure}
  \begin{center}
    \includegraphics[width=160mm]{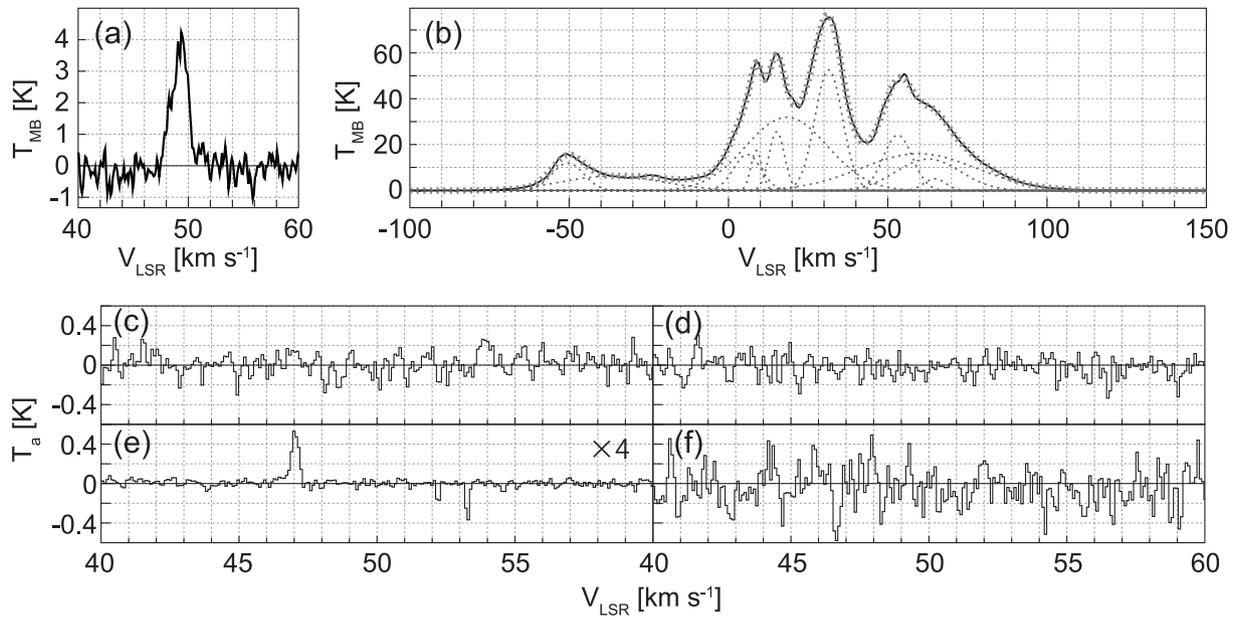}
  \end{center}
  \caption{(a) Smoothed profile in $^{12}$CO($J$=1--0) emission. The smoothed area is equal to the beam size of the Usuda64m telescope. (b) Same as (a) but for H$_{\rm I}$ emission. Thin dotted profiles show the result of multiple Gaussian fitting by ROHSA. The thick dotted profile is the profile that sums up the thin dotted profiles. (c)--(f) Profiles of OH emissions of four $\Delta$-type doubling transition in the ground-state ($J$=3/2 in 2$\Pi$3/2) of (c) 1612 MHz, (d) 1665 MHz, (e) 1667 MHz, and (f) 1720 MHz. The spectrum in (e) is quadrupled vertically.   \label{fig:f6}}
\end{figure}

\begin{figure}
  \begin{center}
    \includegraphics[width=140mm]{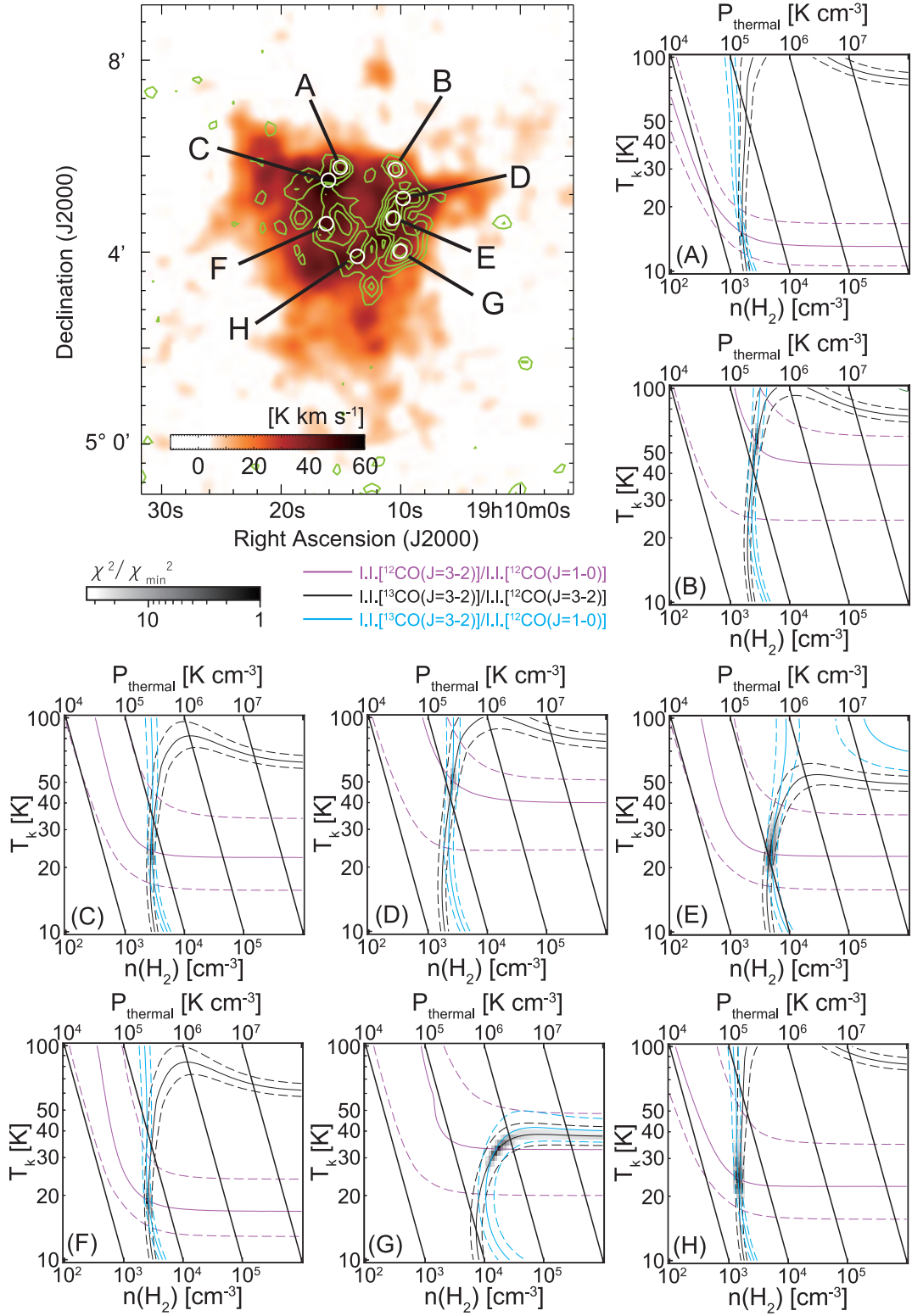}
  \end{center}
  \caption{(Upper-left panel) Integrated intensity map of $^{12}$CO($J$=1--0) emission superposed on positions where the RADEX analysis has been conducted by white circles. Integrated velocity range in the map is 45 km s$^{-1}$ to 55 km s$^{-1}$. The positions are labeled from A to H. Each result of the RADEX analysis is shown in the figure by $n$(H$_2$)-$T_{\rm k}$ plots. Labels are shown in lower-left corner in each $n$(H$_2$)-$T_{\rm k}$ panel. Solid lines in each intensity ratio represent the intensity ratio in each position. The dashed line associated with each solid line represents the 10\% error of the ratio. The gray scale $n$(H$_2$)-$T_{\rm k}$ plane is the ratio of $\chi$$^2$/$\chi$$_{\rm min}$$^2$.   \label{fig:f7}}
\end{figure}

\begin{figure}
  \begin{center}
    \includegraphics[width=140mm]{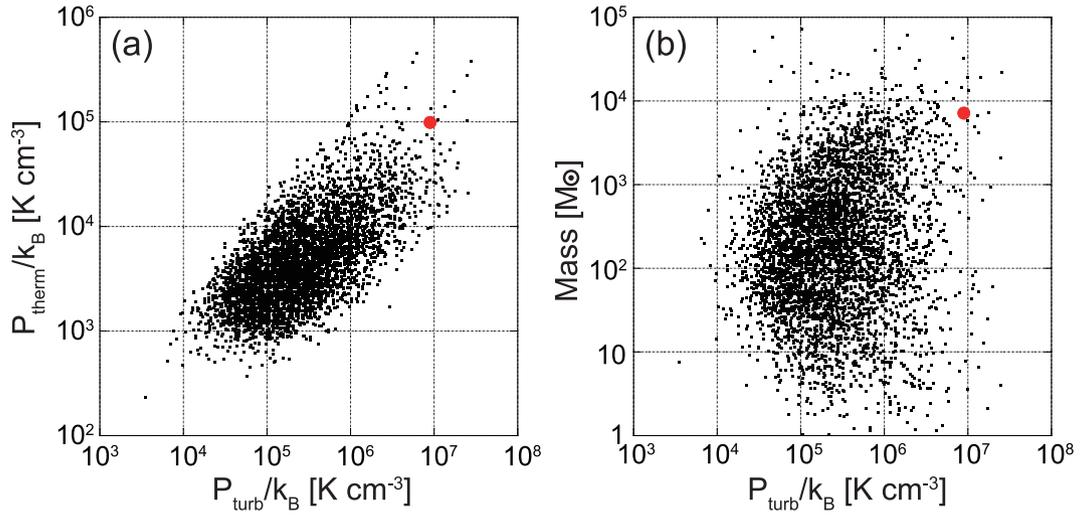}
  \end{center}
  \caption{(a) Plot of the turbulent pressure against the turbulent pressure. Black dots denote the datasets derived by CHIMPS \citep{rig19}. The red dot is that for N4.  (b) Plot of the mass of the molecular cloud with the turbulent pressure. Black and red dots are the same as those in (a).   \label{fig:f8}}
\end{figure}

\begin{figure}
  \begin{center}
    \includegraphics[width=160mm]{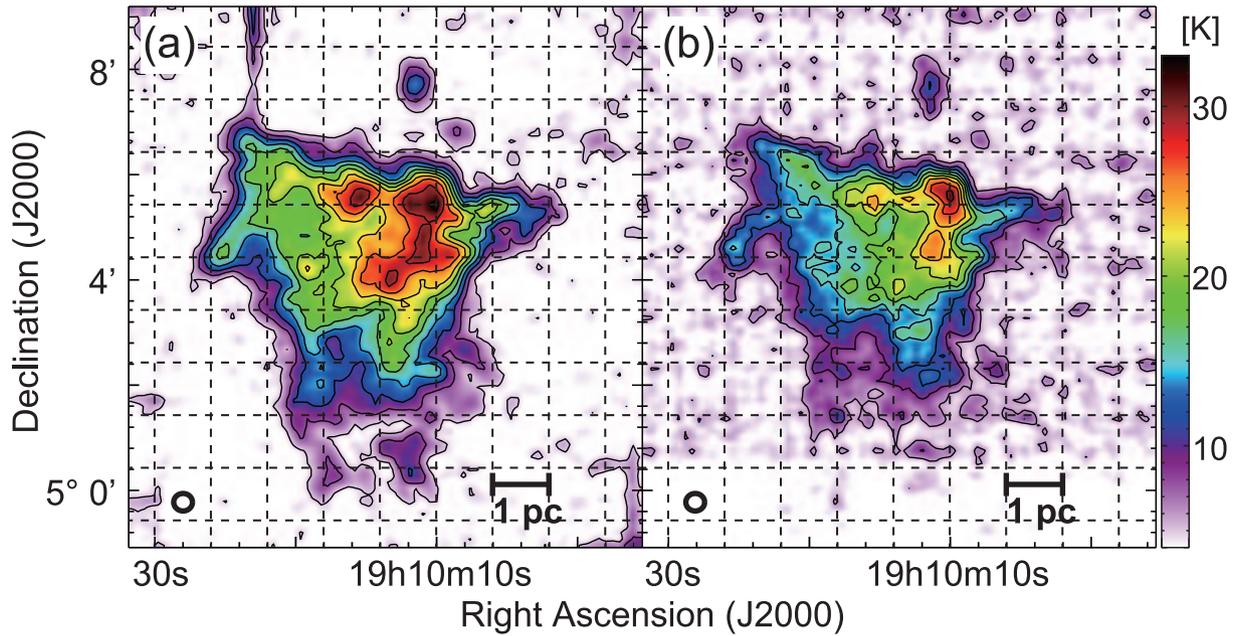}
  \end{center}
  \caption{(a), (b)  Peak T$_{\rm MB}$ distribution in (a) $^{12}$CO($J$=1--0) and (b) $^{12}$CO($J$=3--2) emission, respectively. Dotted grid lines are drawn every 1 pc from the peak T$_{\rm MB}$ position in $^{12}$CO($J$=1--0) emission. The contours in (a) and (b) increase from 5 K to 32 K by a step of 3 K and from 6 K to 30 K by a step of 3 K, respectively. Dotted lines in (b) are the same in (a).  \label{fig:f9}}
\end{figure}

\begin{figure}
  \begin{center}
    \includegraphics[width=160mm]{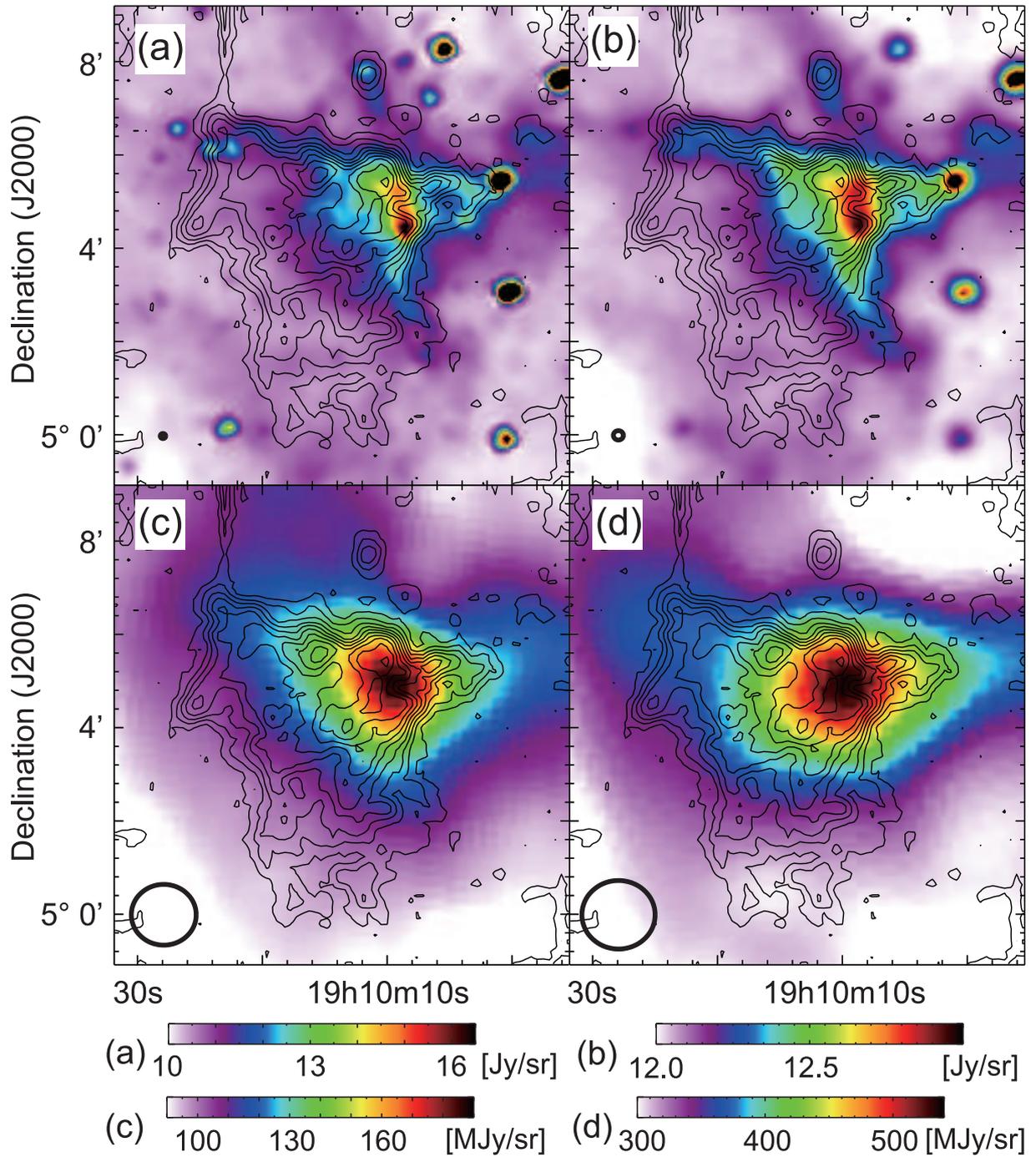}
  \end{center}
  \caption{(a)--(d) Peak $T_{\rm mb}$ distribution in $^{12}$CO($J$=1--0) emission in contours superposed on (a) WISE 12 $\mu$m, (b) WISE 22 $\mu$m, (c) AKARI WideS (90 $\mu$m) and (d) AKARI WideL (140 $\mu$m), respectively. The contour levels increase 5 K to 32 K by a step of 3 K. The bold circle in the lower-left in each panel represents the resolution of the map.   \label{fig:f10}}
\end{figure}

\begin{figure}
  \begin{center}
    \includegraphics[width=160mm]{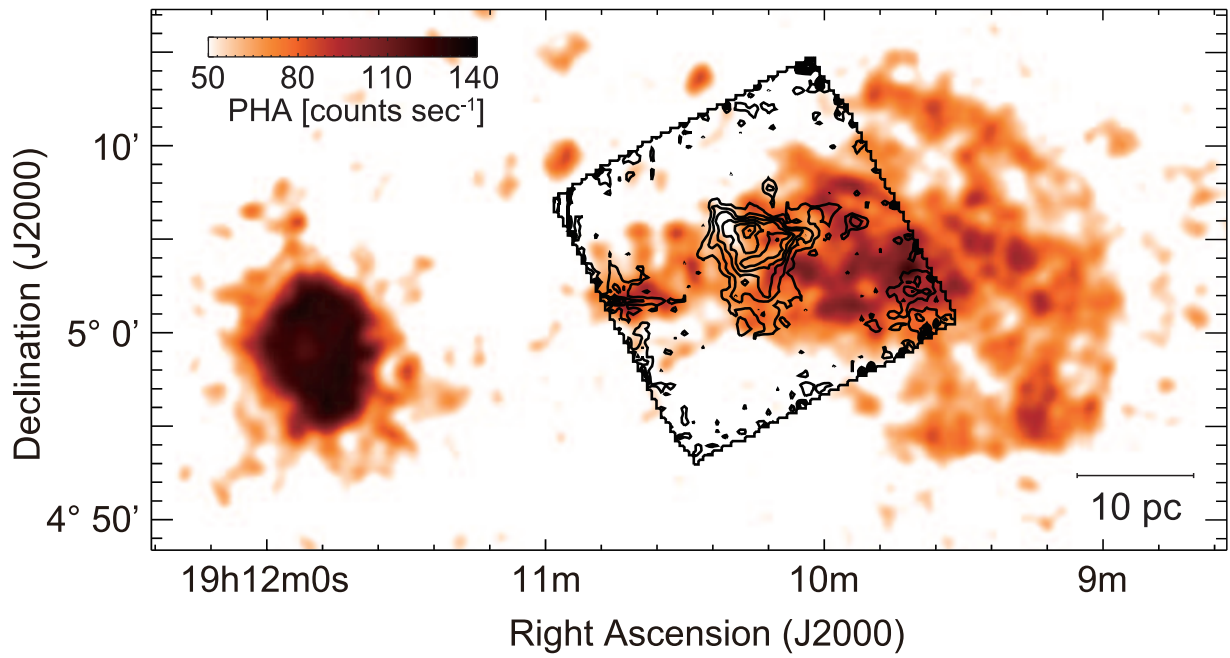}
  \end{center}
  \caption{ROSAT X-ray image superposed on the integrated intensity map of $^{12}$CO($J$=1--0) emission. The contour levels of the CO emission is the same in figure 2(a).  \label{fig:f11}}
\end{figure}

\begin{figure}
  \begin{center}
    \includegraphics[width=160mm]{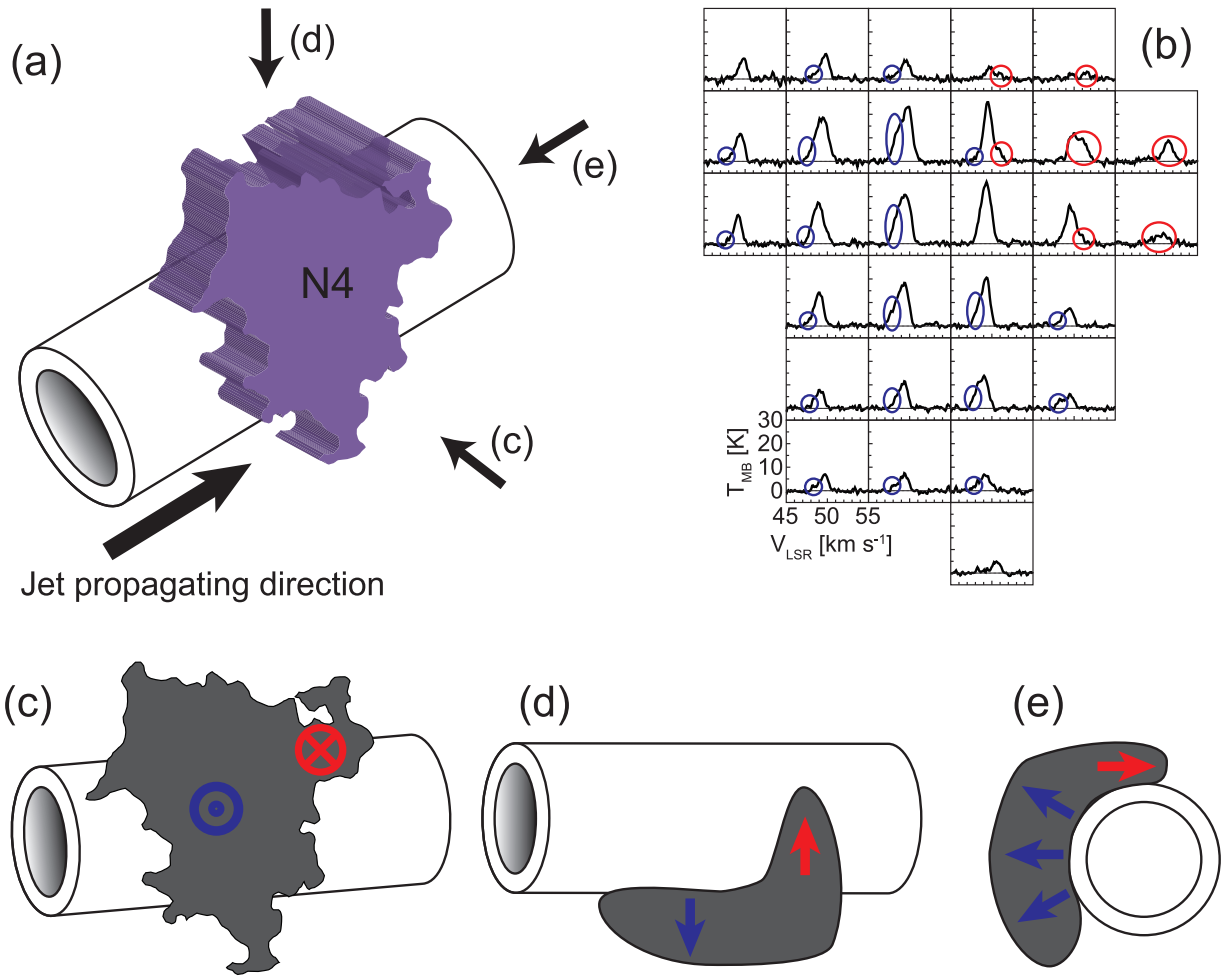}
  \end{center}
  \caption{(a) Schematic view of the interaction of N4 with the jet from SS433.  The direction that the jet propagates is shown by the arrow. The arrows with character (c) to (e) show the direction of the view in figures (c) (front view), (d) (top view), and (e) (side view), respectively. The direction of the arrow (c) shows the direction of the line of sight from us.  (b) the same in figure 6 but only $^{12}$CO($J$=1--0) emission is shown, Blue and red circles show the blue/red-shifted components.  (c), (d), (e)  Front, top, and side views of N4 and the jet. Blue and red symbols show the direction of the motion of the molecular cloud.   \label{fig:f12}}
\end{figure}

\clearpage


\begin{table}{htb}
  \caption{Results of RADEX calculation}
  \begin{tabular}{ccccccccc}
  \hline
  \hline
  Region & \multicolumn{2}{c}{Coordinate} & \multicolumn{3}{c}{Intensity ratio} & $n$(H$_{\rm 2}$) & $T_{\rm k}$ & $P_{\rm thermal}$ \\
   & R.A.(J2000) & Dec.(J2000) & $\frac{^{\rm 12}{\rm CO}(J{\rm =3-2})}{^{\rm 12}{\rm CO}(J{\rm =1-0})}$ & $\frac{^{\rm 13}{\rm CO}(J{\rm =3-2})}{^{\rm 12}{\rm CO}(J{\rm =3-2})}$ & $\frac{^{\rm 13}{\rm CO}(J{\rm =3-2})}{^{\rm 12}{\rm CO}(J{\rm =1-0})}$ & & & \\
   & & & & & & ($\times$ 100 cm$^{-3}$) & (K) & ($\times$ 10$^4$ K cm$^{-3}$) \\
  \hline
  A & 19h10m14.8s & 5$^\circ$5$^\prime$47.7$\arcsec$ & 0.67$\pm{0.04}$ & 0.09$\pm{0.04}$ & 0.06$\pm{0.03}$ & 16$^{\rm +1.9}_{\rm -0}$ & 15$^{\rm +1}_{\rm -1}$ & 2.4$^{\rm +0.4}_{\rm -0.2}$ \\
  B & 19h10m10.3s & 5$^\circ$5$^\prime$44.1$\arcsec$ & 0.95$\pm{0.05}$ & 0.16$\pm{0.03}$ & 0.15$\pm{0.03}$ & 28$^{\rm +3.4}_{\rm -5.8}$ & 56$^{\rm +15}_{\rm -19}$ & 15.8$^{\rm +6.6}_{\rm -7.5}$ \\
  C & 19h10m16.3s & 5$^\circ$5$^\prime$26.1$\arcsec$ & 0.76$\pm{0.02}$ & 0.13$\pm{0.02}$ & 0.10$\pm{0.02}$ & 28$^{\rm +3.4}_{\rm -3.1}$ & 24$^{\rm +8}_{\rm -4}$ & 6.8$^{\rm +3.3}_{\rm -1.8}$ \\
  D & 19h10m10.3s & 5$^\circ$5$^\prime$8.1$\arcsec$ & 0.82$\pm{0.03}$ & 0.11$\pm{0.02}$ & 0.09$\pm{0.02}$ & 25$^{\rm +3.1}_{\rm -2.7}$ & 50$^{\rm +9}_{\rm -10}$ & 12.6$^{\rm +4.0}_{\rm -3.6}$ \\
  E & 19h10m10.7s & 5$^\circ$4$^\prime$46.5$\arcsec$ & 0.80$\pm{0.03}$ & 0.18$\pm{0.03}$ & 0.15$\pm{0.02}$ & 50$^{\rm +20.7}_{\rm -10.3}$ & 24$^{\rm +14}_{\rm -6}$ & 12.0$^{\rm +14.9}_{\rm -4.8}$ \\
  F & 19h10m16.5s & 5$^\circ$5$^\prime$26.1$\arcsec$ & 0.82$\pm{0.03}$ & 0.11$\pm{0.03}$ & 0.09$\pm{0.02}$ & 25$^{\rm +3.1}_{\rm -0}$ & 19$^{\rm +2}_{\rm -1}$ & 4.8$^{\rm +1.1}_{\rm -0.3}$ \\
  G & 19h10m9.7s & 5$^\circ$4$^\prime$3.3$\arcsec$ & 0.91$\pm{0.04}$ & 0.28$\pm{0.03}$ & 0.25$\pm{0.03}$ & 178$^{\rm +9822}_{\rm -65.6}$ & 33$^{\rm +6}_{\rm -7}$ & 58.7$^{\rm +3800}_{\rm -30}$ \\
  H & 19h10m14.6s & 5$^\circ$3$^\prime$48.9$\arcsec$ & 0.73$\pm{0.03}$ & 0.10$\pm{0.03}$ & 0.07$\pm{0.02}$ & 14$^{\rm +3.7}_{\rm -1.5}$ & 24$^{\rm +26}_{\rm -5}$ & 3.4$^{\rm +5.5}_{\rm -1.0}$ \\
  \hline
  \end{tabular}
\end{table}


\begin{ack}
We thank Prof. Masato Tsuboi for helpful comments for improvement and observations in L-band with Usuda 64 m telescope. We also thank Prof. Takeshi Tsuru and Mr. Kazuho Kayama for useful comments and discussion for the interaction of the cloud with X-ray jet. 
This work was supported by JSPS KAKENHI Grant Number 21253003 and 19K03912.
The James Clerk Maxwell Telescope is operated by the East Asian Observatory on behalf of The National Astronomical Observatory of Japan; Academia Sinica Institute of Astronomy and Astrophysics; the Korea Astronomy and Space Science Institute; Center for Astronomical Mega-Science (as well as the National Key R\&D Program of China with No. 2017YFA0402700). Additional funding support is provided by the Science and Technology Facilities Council of the United Kingdom and participating universities and organizations in the United Kingdom and Canada.
The Nobeyama 45 m radio telescope is operated by Nobeyama Radio Observatory, a branch of National Astronomical Observatory of Japan. 
Softwares for K5/VSSP and K5/VSSP32 developed by National Institute of Information and Communication Technology are used to analyze the data taken with USUDA 64m telescope. 
This publication makes use of data products from the Wide-field Infrared Survey Explorer, which is a joint project of the University of California, Los Angeles, and the Jet Propulsion Laboratory/California Institute of Technology, funded by the National Aeronautics and Space Administration.
This research is based on observations with AKARI, a JAXA project with the participation of ESA.
\end{ack}

\end{document}